\documentstyle[twoside, epsf]{article}

\input ibvs2.sty

\begin{document}

\IBVShead{5748}{22 January 2007}

\IBVStitletl{Detection of a Large Flare in FR C\lowercase{nc}}{ (=1RXS J083230.9+154940)}

\IBVSauth{Golovin, A.$^{1, 2, 4}$; Pavlenko, E.$^3$; Kuznyetsova, Yu.$^2$; Krushevska, V.$^2$}

\IBVSinst{Kyiv National Taras Shevchenko University, Kyiv, Ukraine
\\ e-mail: astronom\_2003@mail.ru, astron@mao.kiev.ua}

\IBVSinst{Main Astronomical Observatory of National Academy of Science of Ukraine, Kyiv,
Ukraine}

\IBVSinst{Crimean Astrophysical Observatory, Crimea, Nauchnyj, Ukraine}

\IBVSinst{Visiting astronomer of the Crimean Astrophysical Observatory, Crimea, Nauchnyj,
Ukraine}

\SIMBADobjAlias{FR Cnc}{1RXS J083230.9+154940}

\GCVSobj{FR Cnc}
\IBVStyp{ BY }
\IBVSkey{Photometry, flare}

\IBVSabs{We report detection of an optical flare in the BY}
\IBVSabs{Draconis type star FR Cnc. The flare duration is 41 min, the}
\IBVSabs{amplitude is in the B band 1.02 m. It is the first flare}
\IBVSabs{reported for this object.}

\begintext

FR Cnc (= BD+16\deg1753 = MCC 527 = 1ES 0829+15.9 = 1RXS
J083230.9+154940 = HIP 41889 = GSC 01392-02634 = TYC 1392-2634-1)
($\alpha_{2000}= 08\hr 32\mm 30\fsec5287$ and $\delta_{2000} =
+15\deg49\arcm26\farcs193$) was first mentioned as a probable active
star when it was identified as the optical counterpart  of a soft
X-ray source 1ES 0829+15.9 in the Einstein Slew Survey. It has $V
= 10\fmm43$, spectral type K8V, the X-ray flux is of $\approx
10^{-11}~{\rm erg \cdot s^{-1} \cdot cm}^{-2}$ (Elvis et al., 1992;
Schachter et al., 1996).

It was classified as BY Dra type star (i.e.\ its variability is caused
by rotational modulation of starspots) and given the name FR Cnc
by Kazarovets et al.\ (1999). The presence of Ca\,II H, K and
H$_{\alpha}$ emission lines in the spectra indicates high
chromospheric activity in FR Cnc (Pandey et al., 2002; Pandey,
2003). The other details concerning history of investigation of
this object can be found in Pandey et al.\ (2005)

Flares in FR Cnc were not previously reported.

FR Cnc was observed on 23 November, 2006 quasi-simultaneously
in $B, V, R_{j}, I_{j}$ bands at Crimean Astrophysical Observatory
(Ukraine) by Alex Golovin, using 38-cm Cassegrain telescope, which is
equipped with SBIG ST-9 CCD camera, cooled by a Peltier system to
about $-30~\deg$C{}. The exposure times were 20 s, 13 s, 8 s and
17 s for $B, V, R_{j}, I_{j}$ bands respectively.  Data
reduction was done using ``Maxim DL'' package. Reduction included
bias, dark-frame subtraction and flat field correction using
twilight sky exposures. Since the field of FR Cnc is not crowded,
the technique of aperture photometry was applied to extract the
differential magnitudes. The total number of useful frames was 89
for each band. The brightness of FR Cnc was measured with respect
to GSC 1392-2636 ($\alpha_{2000} = 08\hr 32\mm 23\fsec698;
\delta_{2000} = +15\deg 46\arcm 50\farcs15$), while GSC 01392-02708
($\alpha_{2000} = 08\hr 32\mm 38\fsec2271; \delta_{2000} =
+15\deg 44\arcm 22\farcs095)$ served as a check star. Since the
magnitudes of the comparison star in all bands are not known, here we
present just differential magnitudes.

The data points have a statistical accuracy of $0\fmm 01$ or
better (determined from the difference \emph{check star${}-{}$comparison
star}). To rule out the possibility of observing brightness
variations caused by the comparison star, an independent
photometry of GSC 1392-2636 (comp.\ star) was performed with
respect to the check star (GSC 01392-02708).

\IBVSfig{10cm}{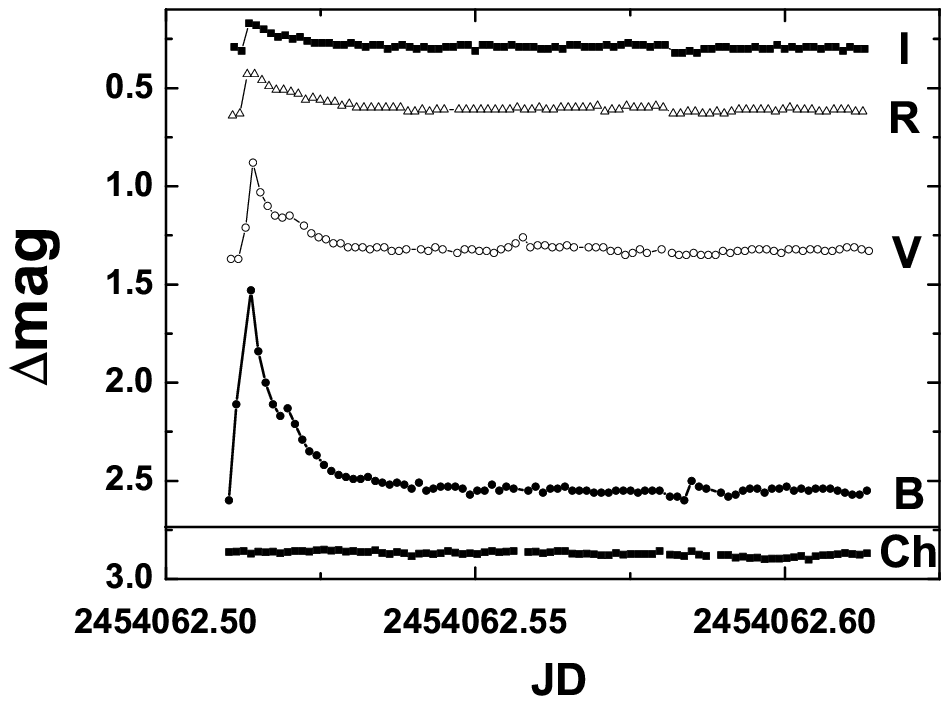}{The flare of FR Cnc: shifted
differential lightcurves in $B, V, R$ and $I$ bands as well as the
difference \emph{check star${}-{}$comparison star} (`Ch' on the
plot)}
\IBVSfigKey{5748-f1.eps}{FR Cnc}{flare light curve}

The flare of FR Cnc was detected on 23 November, 2006 with the
maximum at 00:19 (UT). After the initial rapid flaring, the
brightness of FR Cnc decreased slowly. The time between the flare
began and reached its maximum was about 4 minutes, while the total
duration of the flare was about 41 minutes.

The flare had a maximum amplitude (1\fmm02) in the $B$ band. In
other bands the amplitudes were 0\fmm49, 0\fmm21 and 0\fmm14
for $V, R_{j}$ and $I_{j}$ bands respectively.

Noteworthy, in 8 minutes after the flare's maximum a notable
``spike'' was observed in $B$ and $V$ bands (in other bands the
amplitude was probably too low) during the brightness decline.
Remarkable, that FR Cnc remained to be about 0\fmm05 brighter
for at least an hour after the flare began comparing with
brightness before flare.

Following the idea, described at Kozhevnikova et al.\ (2006), we calculated the
intensity of the flare and the \emph{absolute} energy output. The relative
intensity of the flare was determined via the following relation:
$\frac{I_f}{I_0} =(\frac{I_{0} +I_{f}}{I_0})-1$, where $I_{0}+I_{f}$ is the
intensity of the object, integrated over the duration of the flare, $I_0$ is the
intensity of the star in quiescent level in one of the bands (corrected to the
flare duration). For calculation of the \emph{absolute} energy output, we assume
for FR Cnc's quiescent level the following magnitude and colour indices: $V =
10.43, B-V = 1.35, V-R = 1.15, V-I = 1.93$. We used $30.24 \pm 2.03$ mas
parallax (Perryman et al., 1997) that imply distance $33 \pm 2$ pc.

Similar calculations of the flare intensity and energy output were also done by
Moffett (1973) and by Panov et al.\ (2000).

\begin{table}
\begin{center}
{\normalsize Table 1. Flare properties} \vskip 3mm
\begin{tabular}{cccc}
\hline Band  & Amplitude [mag] & Flare flux/quiescent flux [\%] & Flare energy [erg / \AA]\\
\hline
$B$ & 1.02 & 38.63 & $1.73\times10^{31}$ \\ 
$V$ & 0.49 & 14.05 & $1.14\times10^{31}$ \\
$R$ & 0.21 & \phantom{0} 8.25 & $0.89\times10^{31}$  \\ 
$I$ & 0.14 &  2.9  & $0.29\times10^{31}$ \\
\hline
\end{tabular}
\end{center}
\end{table}

So, we get the values listed in Table~1. Fig.~1 shows differential lightcurves
in $B, V, R_{j}$ and $I_{j}$ bands of FR Cnc during our observations on 23
November, 2006.

However, the observed rotational period ($0.8267 \pm 0.0004$ from
Pandey et al., 2005) is unusually short for such type of stars, which 
implies that this star should manifest strong flaring
activity (see Dorren et al., 1994). We detected a flare of FR Cnc 
for the first time. Further monitoring of this object is highly
desirable.

\bigskip

\textbf{Acknowledgements:} First of all, it is a great pleasure
for the authors to express here sincere thanks to Galvez Mari Cruz
(Depto.\ Astrofisica, Universidad Complutense de Madrid, Madrid,
Spain) for pointing our interest to this object. Authors are very
grateful to R. Gershberg, A. Kozhevnikova and I. Alekseev for
valuable comments. Alex Golovin indebted to Jevgeniy Kachalin for
his great help with preparation this manuscript and for the
proof-reading. It is a great pleasure for Alex Golovin to express
personal thankfulness to Maksim Andreev (Terskol Branch of the RAS
Institute of Astronomy, Terskol, Russia) for useful discussions
and suggestions during preparation of this paper. 

\references

Dorren, J.D., Guinan, E.F., Dewarf, L.E., 1994, {\it ASPCS}, {\bf 64}, 
399 (Cool stars, stellar systems, and the Sun, ed.\ J.-P. Caillault)

Elvis, M., et al., 1992, {\it ApJS}, {\bf 80}, 257

Kazarovets, A.V., et al., 1999, {\it IBVS}, No.\ 4659

Kozhevnikova, A.V., Alekseev, I.Yu., et al., 2006, {\it IBVS}, No.\ 5723

Moffett T.J., 1973, {\it Mon.\ Not.\ R. Astr.\ Soc.}, {\bf 164}, 11

Pandey, J.C., et al., 2002, {\it IBVS}, No.\ 5351

Pandey, J.C., 2003, {\it Bull.\ of the Astron.\ Soc.\ of India}, {\bf 31}, 329

Pandey, J.C., et al., 2005, {\it AJ}, {\bf 130}, 1231

Panov, K., Goranova, Yu., Genkov, V., 2000, {\it IBVS}, No.\ 4917

Perryman, M.A.C., et al., 1997, {\it A\&A}, {\bf 323}, L49

Schachter, J.F., et al., 1996, {\it ApJ}, {\bf 463}, 747

\endreferences

\end{document}